\documentclass[review]{elsarticle}

\usepackage{hyperref}
\usepackage{array}
\usepackage{appendix}
\usepackage{mathtools}
\usepackage[table]{xcolor}
\usepackage{caption}
\usepackage{multirow}
\usepackage{lscape}
\usepackage{graphicx}
\usepackage{siunitx}
\usepackage{ntheorem}
\theoremseparator{:}
\newtheorem{hyp}{Hypothesis}

\definecolor{DeepRed}{HTML}{990000}
\definecolor{DarkSalmon}{HTML}{ea9999}
\definecolor{Salmon}{HTML}{f4cccc}

\definecolor{DeepPurple}{HTML}{674ea7}
\definecolor{DarkLiliac}{HTML}{b4a7d6}
\definecolor{Liliac}{HTML}{d9d2e9}

\definecolor{DeepGreen}{HTML}{6aa84f}
\definecolor{DarkLime}{HTML}{b6d7a8}
\definecolor{LightGreen}{HTML}{d9ead3}

\definecolor{DarkBlue}{HTML}{6d9eeb}
\definecolor{LightBlue}{HTML}{c9daf8}

\definecolor{Plum}{HTML}{a64d79}
\definecolor{LightPlum}{HTML}{c27ba0}
\definecolor{SweetPlum}{HTML}{dbaac4}
\definecolor{Mustard}{HTML}{ffd966}
\definecolor{DeepSalmon}{HTML}{e06666}

\makeatletter
\newcommand*{\@rowstyle}{}
\newcommand*{\rowstyle}[1]{
  \gdef\@rowstyle{#1}%
  \@rowstyle\ignorespaces%
}
\newcolumntype{=}{
  >{\gdef\@rowstyle{}}%
}
\newcolumntype{+}{
  >{\@rowstyle}%
}
\makeatother

\journal{Information Processing \& Management}

\biboptions{authoryear}

\begin{document}

\begin{frontmatter}

\title{Measuring the influence of mere exposure effect of TV commercial adverts on purchase behavior based on machine learning prediction models}

\author[gidai]{Elisa Claire Alem\'an Carre\'on
\corref{mycorrespondingauthor}}
\ead{s153400@stn.nagaokaut.ac.jp}

\author[gidai]{Hirofumi Nonaka}
\ead{nonaka@kjs.nagaokaut.ac.jp}

\author[gidai]{Asahi Hentona}
\ead{s173348@stn.nagaokaut.ac.jp}

\author[gidai]{Hirochika Yamashiro}
\ead{s173358@stn.nagaokaut.ac.jp}

\address[gidai]{Nagaoka University of Technology, Nagaoka, Japan}

\cortext[mycorrespondingauthor]{
Corresponding author
}

\begin{abstract}
Since its introduction, television has been the main channel of investment for advertisements in order to influence customers purchase behavior. Many have attributed the mere exposure effect as the source of influence in purchase intention and purchase decision; however, most of the studies of television advertisement effects are not only outdated, but their sample size is questionable and their environments do not reflect reality. With the advent of the internet, social media and new information technologies, many recent studies focus on the effects of online advertisement, meanwhile the investment in television advertisement still has not declined. In response to this, we applied machine learning algorithms SVM and XGBoost, as well as Logistic Regression, to construct a number of prediction models based on at-home advertisement exposure time and demographic data, examining the predictability of Actual Purchase and Purchase Intention behaviors of 3000 customers across 36 different products during the span of 3 months. If we were able to predict purchase behaviors with models based on exposure time more reliably than with models based on demographic data, the obvious strategy for businesses would be to increase the number of adverts. On the other hand, if models based on exposure time had unreliable predictability in contrast to models based on demographic data, doubts would surface about the effectiveness of the hard investment in television advertising. Based on our results, we found that models based on advert exposure time were consistently low in their predictability in comparison with models based on demographic data only, and with models based on both demographic data and exposure time data. We also found that there was not a statistically significant difference between these last two kinds of models. This suggests that advert exposure time has little to no effect in the short-term in increasing positive actual purchase behavior.

\paragraph{Highlights}
\begin{itemize}
\setlength\itemsep{-0.5em}
\item Models based on exposure time to television adverts have significantly lower predictability of actual purchase in comparison with models using demographic data.
\item Actual Purchase behavior predictability was not significantly different in models including both demographic data and exposure time data as opposed to those with only demographic data.
\item Results suggest that advert exposure time has little to no effect in the short-term in increasing positive actual purchase behavior.
\end{itemize}

\end{abstract}

\begin{keyword}
Television Adverts\sep
Purchase Behavior\sep
SVM\sep
XGBoost\sep
Machine Learning
\end{keyword}

\end{frontmatter}

\section{Introduction}
\label{intro}

It is generally thought that in order for companies to increase sales, they must somehow increase the purchase intention of their potential customers \cite[][]{armstrong,morwitz}. Historically this has been approached through many channels, but since the successful introduction of the television to the general public, it has been largely attempted via television commercial advertisements, and many companies invest heavily on these efforts. However, most studies to prove the effectiveness of these advertisements have been conducted on small sample groups, usually introducing a customer to a commercial advertisement and measuring their intentions to purchase a product before and after watching the advertisement with a survey \cite[e.g.][]{khuong}. Studies on the predictability of purchase behavior from purchase intention data have pointed out that many of these analyses have very different results \cite[][]{morwitz,sun,newberry}, presumably because of small and non-representative samples, and controlled environments that do not reflect reality. 

With the advent of Big Data and new methodologies in the field of information technology, there is a new and improved lens for advertisement research on the effects it can have on people outside controlled environments; however, its focus is mostly on similarly new advertisement online and in social media \cite[][]{shareef-ma,gonzalez,ramaboa,wu}, leaving behind the study of more traditional advertisement which has not declined in use since the increase of online advertisement. In response to this lack of current research in the field of television advertisement, we propose a machine learning approach to this problem, with a large database of the household television usage timelines of surveyed individuals and their answers regarding recent purchase intentions and actual purchase recalls at two points in time separated by 3 months, provided by the Nomura Research Institute, Ltd. 

Now, following the traditional train of thought of the effects of mere exposure \cite[][]{zajonc}, we propose collecting the accumulated number of seconds that a user has viewed a commercial advert related to a certain product and observe its effects on the users. With this data we propose training a number of models to predict the purchase intention and purchase recall of users based on the amount of accumulated seconds of being exposed to the advertisement for the related product in the survey, and then compare it to models that use demographic data of the users as a control. We propose to do this by unit of product, to observe the difference in marketing success from product to product, and by unit of user, to observe the rate of population that was potentially influenced by advertisements. This introduces both granularity, as we are using precise television viewing time and observing effects over time, and the potential of generalizing our prediction model to unknown new users or products after training.

\section{Research objective}
\label{resobj}

The objective of this study is to provide an updated methodology and a larger scale database to measure the mere exposure effect and perceptual fluency effect of television adverts on purchasing behavior. For a long time, psychology based studies have been widely performed on small groups of people in very controlled environments that do not reflect customers in real life accurately, and they have been traditionally thought effective without criticism. We aim to measure the predictability in purchase behavior based on the time spent exposed to adverts of specific products in household televisions during the duration of 3 months, then compare it to the predictability in purchase behavior when using demographic data to provide a clearer answer to whether the heavy investment into TV advertising is actually having an effect on customers to purchase more. As control, we will also measure the predictability in purchase behavior based on demographic data and combining the two sources of data. In the case that the predictability is high enough compared to models that don't include exposure time, this methodology could be used as a measure for future sales. On the other hand, a low predictability in comparison to the control would create doubts on whether the mere exposure to advertisements on television is being effective.

\section{Related work}
\label{related}

In previous research, there have been attempts to analyze the effects of adverts via mere exposure \cite[][]{hekkert}, and many studies have replicated the original experiment by Zajonc unrelated to adverts in the field of psychology \cite[][]{huang,dechene}. Now, in addition to the focus on the mere exposure effect, there have been attempts to measure the effects of advertisements on brand recognition and perception fluency \cite[][]{fang}, as well as its effects on the perception of the product \cite[][]{gmuer}. Fluency is defined as the level of ease or difficulty with which external information is processed \cite[][]{schwarz}. Previously it has been proven that it can produce bias, and it has been shown to affect the judgement of truth \cite[][]{silva}. For a long time, the perceptual fluency model has stated that repeated exposure leads to a more readily accessibility of the target brand in memory, which in turn must have an effect on the ability to recognize a brand in the future \cite[e.g.][]{jacoby}. Most of the older research had arrived to a consensus that there is a positive influence \cite[][]{reber, seamon}. More recent research, however, explores further whether these effects in memory are strictly related to positive emotional judgment on the brands or if they can also imply negative judgements based on the main objective of a product \cite[][]{lee-a}. 

Research of the direct effects of television advertisement has also been attempted. One study focuses on child obesity by using weight measurements \cite[][]{boyland}. An even more direct approach has been made in another study which has used brain imaging in order to explore the short-term and long-term memory effects of TV commercials \cite[][]{rossiter}. It should be noted that, as is to be expected in a brain imaging experiment, the participants observed the advert directly and more consciously than in mere exposure experiments. 

Now, two of the main issues with these studies and others in television advertisement effects are that, not only is the size of the samples in these experiments questionably small, but the environment is limited in that it becomes extremely controlled, to the point where it doesn't reflect the reality of customers watching daily television in their homes and making purchases anymore, and the observations environment itself could affect the results. 

In order to solve this limitation, our research is based in data science analysis methodology, such as machine learning algorithms trained from large samples of data. Current big data analysis on advertising is mostly focused on online advertisements \cite[][]{wu, stitelman}, where, with the advance of current technology, a user is exposed to adverts placed near to the content they are currently consuming which are specifically targeting their interests \cite[][]{perlich,schwartz}, catching their attention (which is no longer mere exposure, but direct interaction), or a user is incentivized to watch an advertisement by blocking completely the content they were consuming until the advertisement is finished playing on screen. Most of the research in this area is focused on new ways to create online advertisements in social media \cite[][]{shareef-mb} and suggestions or recommendations targeted to a user's interests \cite[e.g.][]{jansen,zhang,kannan,choi} reducing the need of mere exposure advertisement while online. In addition to this, some studies have focused on testing the effects of online advertisement on customers \cite[][]{alawan,lee-j,shareef-mc}.

While these new technologies made possible the analysis of online advertisement and social media, the focus has shifted and there is no research using these technologies to test the effectiveness of the mere exposure effect based advertisements which are still in use in other traditional means, such as billboards, or as we analyze in our study, television advertisement. Our study is unique in that, using data from television advertisement in household environments and not online ads, we apply data science methodology to explore with a larger sample and a household environment, if there is an effect caused by mere exposure advertisement, and to what extent this effect happens. Our study is also unique in that comparing the results of prediction models based on exposure time to those using demographic data used in previous literature, we can determine if there is an effect caused by exposure, or if purchase behavior is better decided by other external factors, such as income of each individual and their marital and parental statuses.

\section{Methodology}
\label{method}

As explained above, our approach is to train machine learning models based on the number of seconds of advertisement exposure and demographic data, to predict the effect on the customers purchase decisions measure their predictability. A high predictability based on exposure time would be useful for measuring and predicting sales in any industry. On the other hand, a low predictability in comparison to that of demographic data models would create doubts that the current advertisement based on mere exposure is effective.

Our proposed method is explained in detail in the following sections.

\subsection{Experiment design overview}
\label{exp_design}

First we will explain the general design of the experiments. Each experiment consists in creating a prediction model based on a dataset comprised of input features and previously known output labels. After the model is trained, it is able to make predictions of new output labels of unknown data if given new input values. In this study, we created many models by variating the training input features and output labels and compare their results. For the input data, we prepared datasets based on advertisement viewing time and demographic data. For the prediction targets, we prepared datasets for purchase intention and actual purchase behaviors. We also measured the predictability of each purchase behavior target either by unit of product, to observe the difference in marketing success from product to product, and by unit of user, to observe the rate of population that was potentially influenced by advertisements. Finally we utilized 3 different prediction models, Support Vector Machine \cite[][]{svm}, XGBoost \cite[][]{xgboost} and Logistic Regression \cite[][]{logit} in order to compare performance. These variations for each experiment are shown in Table \ref{tab:exp_var} and each item will be explained in detail in the following sections. It is important to note that Purchase Intention is described to be used in the input vectors in our experiments in Table \ref{tab:exp_var}, but this was of course removed for the experiments in which it was the Prediction Target to avoid redundancies.

\begin{table}[]  \centering
\caption{Experiment variations.}\label{tab:exp_var}
\begin{tabular}{|m{12em}|m{18em}|} \arrayrulecolor{white}\hline
\rowcolor{DeepPurple}
\color{white}\textbf{Experiment Contents} & \color{white}\textbf{Variations} \\ \hline
\cellcolor{DarkLiliac}\begin{tabular}[c]{@{}l@{}}\textbf{Prediction} \\ \textbf{Model Bases}\end{tabular}
        & \cellcolor{Liliac}\begin{tabular}[c]{@{}l@{}}
        $\bullet$ Product Based Models\\ 
        $\bullet$ User Based Models\end{tabular} \\ \hline
\cellcolor{DarkLiliac}\textbf{Prediction Targets}
        & \cellcolor{Liliac}\begin{tabular}[c]{@{}l@{}}
        $\bullet$ Purchase Intention\\ 
        $\bullet$ Actual Purchase\end{tabular} \\ \hline
\cellcolor{DarkLiliac}\textbf{Input Data Variants}
        & \cellcolor{Liliac}\begin{tabular}[c]{@{}l@{}}
        $\bullet$ Advert Viewing Time\\ 
        $\bullet$ Advert Viewing Time, Demographics, \\ (and Purchase Intention)\\ 
        $\bullet$ Demographics (and Purchase Intention)\end{tabular} \\ \hline
\cellcolor{DarkLiliac}\textbf{Prediction Models}
        & \cellcolor{Liliac}\begin{tabular}[c]{@{}l@{}}
        $\bullet$ Support Vector Machine\\ 
        $\bullet$ XGBoost\\ 
        $\bullet$ Logistic Regression \end{tabular} \\ \hline
\end{tabular}
\end{table}

\subsection{Prediction model bases}
\label{pred_model_bases}

In this study, we measured the predictability of each purchase behavior target either by unit of product, to observe the difference in marketing success from product to product, and by unit of user, to observe the rate of population that was potentially influenced by advertisements. After extracting the commercial advert viewing data using these parameters from the 3000 users that answered the survey, which includes purchase behavior questions from 200 products at two different points in time, only 36 products from those in the survey were linked to commercial adverts that were actually viewed by those same users. Thusly, we performed our experiments using the viewing data of 3000 users for these 36 products in the configurations explained before in Table \ref{tab:modelbases}.

\begin{table} \centering
\caption{Prediction model bases.}\label{tab:modelbases}
\rowcolors{2}{DarkLime}{LightGreen}
\begin{tabular}{|>{\raggedright\arraybackslash}=m{10em}|+m{22em}|} \arrayrulecolor{white}\hline
\rowcolor{DeepGreen}\rowstyle{\color{white}\bfseries}
Prediction Model Base & Description \\ \hline
\begin{tabular}[c]{@{}l@{}}\textbf{Product Based} \\ \textbf{Prediction Models} \end{tabular} 
    & For each product from 36 available in the survey, data from 3000 users was collected and paired with their labels. \\ \hline
\begin{tabular}[c]{@{}l@{}}\textbf{User Based} \\ \textbf{Prediction Models} \end{tabular} 
    & For each user from 3000 available, data corresponding to the 36 products available in the survey was collected and paired with their labels. \\ \hline
\end{tabular}
\end{table}

\subsection{Prediction targets}
\label{pred_targets}

\subsubsection{Purchase Intention and Actual Purchase}
\label{pi_and_ap}

From the survey data provided by Nomura Research Institute Ltd., we can examine 3000 customer samples, of which we can extract the Purchase Intention and Actual Purchase answers at two points in time, one in January 2017, and another in March 2017, for 200 different products, 36 of which had advertisemnts in the same time period. Each time, the surveys inquire the customer if they have recently had an intention or desire to purchase a certain product (regardless of action on this desire), which corresponds to Purchase Intention; likewise, it inquires if they have recently had purchased a product, corresponding to the Actual Purchase element. We will inspect the effect of adverts on these two elements of a customer's purchase decisions and observe their change with time on the span of three months.

\subsubsection{Prediction target data categorization}
\label{target_data_cat}

In order to explore the different effects commercial adverts may have on the purchase decisions of customers based on their answers from two different points in time, we have labeled each user in regard to each product with 6 categories (from 0 to 5), describing several patterns of behavior. For example, let's examine customers who answered they had purchased a product in January and then not in March, corresponding to category 0, in comparison to customers who purchased the product in March, corresponding to category 4: It is possible that, had category 0 customers were exposed to adverts in greater quantity than other users who still purchased the product and weren't exposed to as many adverts on the span of 3 months, this could mean that the advert was at least not effective, or in a worse scenario, off-putting. On the other hand, if the amount of advert exposure was minimal with category 0 customers and at the same time, customers in category 4 who actually recall having purchased the product in the March survey had been exposed to a large amount of adverts, it would prove to be an effective commercial advert campaign.

Although our approach for analysis is different, the above is a simple example of the importance of this distinction between behavior categories. The six categories for each element are explained in detail in Table \ref{tab:categories_ap} and Table \ref{tab:categories_pi}.

\begin{table}[htp] \centering
\caption{Category definition for Actual Purchase element.}\label{tab:categories_ap}
\rowcolors{2}{DarkSalmon}{Salmon}
\begin{tabular}{|=c|>{\centering\arraybackslash}+m{13em}|>{\centering\arraybackslash}+m{13em}|}\arrayrulecolor{white}\hline
\rowcolor{DeepRed}
\rowstyle{\color{white}\bfseries}
Category & January Actual Purchase & March Actual Purchase \\ \hline
0 & Yes & No \\ \hline
1 & No & No \\ \hline
2 & No & Yes \\ \hline
3 & Yes & Yes \\ \hline
4 & Yes/No & Yes \\ \hline
5 & Yes/No & No \\ \hline
\end{tabular}
\end{table}

\begin{table}[htp] \centering
\caption{Category definition for Purchase Intention element.}\label{tab:categories_pi}
\rowcolors{2}{DarkLiliac}{Liliac}
\begin{tabular}{|=c|>{\centering\arraybackslash}+m{13em}|>{\centering\arraybackslash}+m{13em}|}\arrayrulecolor{white}\hline
\rowcolor{DeepPurple}
\rowstyle{\color{white}\bfseries}
Category & January Purchase Intention & March Purchase Intention \\ \hline
0 & Yes & No \\ \hline
1 & No & No \\ \hline
2 & No & Yes \\ \hline
3 & Yes & Yes \\ \hline
4 & Yes/No & Yes \\ \hline
5 & Yes/No & No \\ \hline
\end{tabular}
\end{table}

\subsection{Input data}
\label{input_data}

\subsubsection{Advert viewing time}
\label{advert_viewtime}

We extracted the viewing time for adverts of each product for each customer from the household television viewing data collected and provided by Nomura Research Institute Ltd. Now the data provided tells us if a user had their personal television turned on at the moment of a certain show. Using the information provided of which commercial advert was shown during which television show and how long they lasted, we extracted the number of accumulated seconds a user had the television on for the adverts of each product, and organized them into different weekdays. We called this the Weekday data configuration. For comparison, in a different model, we separated each weekday into two time slots. We did this to further analyze whether the time period regularly described as "Primetime" (19:00 to 23:00) had any different influence than other time slots. We called this the Weekday Time Slot data configuration. We show the detailed features in \ref{appendix:inputs} in Table \ref{tab:viewtimes}.

\subsubsection{Demographic data}
\label{demographics}

In order to perform control experiments, in which the prediction is either aided by, or designed only to be based on external factors from the advert exposure time, we performed experiments using the demographic information of each user collected at the time of the survey by Nomura Research Institute, Ltd. We used the age, sex, marital status, parental status and income bracket reported by each user. The answers and consequently the vector features are shown in detail in \ref{appendix:inputs} in Table \ref{tab:demographics}.

\subsection{Prediction models}
\label{pred_models}

In this study we chose 3 prediction models: Support Vector Machine, XGBoost and Logistic Regression. SVM and XGBoost are considered well performing supervised machine learning models in the machine learning field considering the size of the data available for this study. Logistic Regression is a statistical model commonly used for binary prediction that is also appropriate for the size of our data. We explain each of those models in more detail in the following sections.

\subsubsection{Support Vector Machine}
\label{svm}

Support Vector Machines (later abbreviated SVM) are supervised machine learning models used in regression and classification problems \cite[][]{svm}. Supervised learning meaning that the model trains on previously labeled data, and establishes a way to match the labels as accurately as possible for new unlabeled data to be analyzed. In a binary classification problem, also called a Support Vector Classifier (SVC), previously established binary labels are matched with a p-dimensional vector of input data. Each column or dimension in the vector expresses a feature in the input data, and each row of the vector is a different data point. After each data point is matched with a label, an SVM uses an algorithm to determine a (p-1)-dimensional hyperplane that separates the p-dimensional space in a way that minimizes error in classification, by maximizing the distance between the hyperplane and the nearest point in either classification. In our study we used the linear kernel for our SVC.

\subsubsection{XGBoost}
\label{xgboost}

Originally started as a research project by Tianqi Chen \cite[][]{xgboost}, XGBoost is an improved and optimized application of a Gradient Boosting Machine, or GBM, also called gradient tree boosting, or gradient boosted regression tree. A Gradient Boosted Regression Tree (GBRT) works by building an ensemble model from several weak learning machines which are just above random guessing in accuracy, in this case using Decision Trees. The misclassified results from these weak predictions are then weighted and added to a final strong learning machine. This process iteratively optimizes the misclassification cost in a functional gradient descent so that the final learning machine focuses on important factors from the training data for a stronger prediction model.

\subsubsection{Logistic regression}
\label{logit_regression}

The logistic model \cite[][]{logit} uses a logistic function to model a binary dependent variable. It is a form of regression in which the probability of the dependant variable being one of two possible values (0 or 1) is estimated from the independent variables. 

\subsection{Model evaluation metrics}
\label{model_evaluation}

In order to measure the effectiveness of the training process and data, we performed what is called a K-fold cross validation. This means that after randomly shuffling and splitting our training data in k equal parts, k-1 of those parts are used for training, while the remaining one part is used in validation. Using the trained models, a prediction is made, and it is decided if such a prediction is correct or not, and counted and grouped as a True Positive, True Negative, False Positive or False Negative prediction. This is explained in Table \ref{tab:preds}.

\begin{table}[htp] \centering
\caption{Prediction outcomes.}\label{tab:preds}
\begin{tabular}{|>{\centering\arraybackslash}m{7em}|>{\centering\arraybackslash}m{7em}|>{\centering\arraybackslash}m{7em}|} \arrayrulecolor{white}\hline 
\multicolumn{1}{c|}{} & \cellcolor{Mustard}\textbf{Prediction is Correct} & \cellcolor{LightPlum}\textbf{Prediction is Incorrect} \\ \hline
\cellcolor{DarkBlue}\textbf{Prediction is Positive} & \cellcolor{DeepGreen}True Positive & \cellcolor{DeepPurple}False Positive \\ \hline
\cellcolor{DeepSalmon}\textbf{Prediction is Negative} & \cellcolor{orange}True Negative & \cellcolor{Plum}False Negative \\ \hline
\end{tabular}
\end{table}

Measures of accuracy are determined from these prediction outcomes. This process is then repeated \(k\) times and the measures taken are averaged. In this study we used the \(F_{1}\) score, which measure is a harmonic mean between precision and recall. Precision, described in formula (\ref{eq:6}), lets us observe the rate of correct positive predictions from all the positive predictions, while Recall, detailed in formula (\ref{eq:7}), observes the rate of correct positive predictions from the total of actual positive data. The \(F_{1}\) score in formula (\ref{eq:8}) then can only be high when both of these measures are high simultaneously, and will lower substantially if they are not consistent. We use this score as it allows us to avoid overlooking data while maintaining accurate predictions.

\begin{equation}\label{eq:6}
Precision = \frac{True Positives}{True Positives + False Positives}
\end{equation}

\begin{equation}\label{eq:7}
Recall = \frac{True Positives}{True Positives + False Negatives}
\end{equation}

\begin{equation}\label{eq:8}
F_{1} = 2  \frac{Precision * Recall}{Precision + Recall}
\end{equation}

\section{Experiments}
\label{experiments}

\subsection{Model training}
\label{model_training}

As explained in section \ref{exp_design}, we designed the experiment by training variations of models depending on the input and output values. The combinations of configurations for the input data shown were explained in Table \ref{tab:exp_var}, and they give us a total of \num[group-separator={,}]{30360} possible inputs for experiment variations. The possible targets explained in Tables \ref{tab:categories_ap} and \ref{tab:categories_pi} give us a total of 12 possible prediction targets. Together, we performed a total of \num[group-separator={,}]{364320} experiments per prediction model. Since we used 3 kinds of prediction models (SVM, XGBoost and Logistic Regression), we performed a total of \num[group-separator={,}]{1092960} experiments in this study.


\subsection{Experiment parameters}
\label{params}

Each prediction model, SVM, XGBoost and the Logistic Regression function can have different parameters when fitting the data to the model. In this study the parameters were chosen broadly to make a general approach (not very specialized) to all the different configurations of the experiment that could take place. Because of the number of experiments explained in section \ref{model_training}, to choose parameters in a specific manner could unbalance one experiment in favor of the other. As such, we chose simple parameters that can apply to many cases.

The SVM experiments were performed with a linear kernel and a \(C\) value of 1. The \(C\) parameter allows for misclassification in exchange of a larger margin at small values, and it becomes stricter for larger values, perhaps causing overfitting if large enough. The XGBoost experiments were performed with a learning rate of 0.1, a maximum tree depth of 3, and 100 estimators. The Logistic Regression experiments were performed with unit weight per individual sample. A 5-Fold cross validation was performed for all of the models.

\section{Results}
\label{results}

Because of the large number of experiments performed in this study, we analyze the average performances for different variations of the model input and prediction output. In order to compare the performance across different variations, we performed \(t\)-tests and examined the p-values for statistical significance. The average performance results are detailed in section \ref{av_res}. The \(t\)-test comparisons are shown in section \ref{h_res}.

\subsection{Prediction score averages}
\label{av_res}

The \(F_{1}\) scores for the SVM product based model for all 36 products were averaged for each variation of the experiment. The results are shown in Table \ref{tab:svm_product_av_res}. The average \(F_{1}\) scores for the SVM user based model for all 3000 users are shown in Table \ref{tab:svm_user_av_res}.

Similarly, the XGBoost product based models average \(F_{1}\) scores are shown in Table \ref{tab:xgboost_product_av_res}, and the user based models average \(F_{1}\) scores are shown in Table \ref{tab:xgboost_user_av_res}.

Lastly, the Logistic Regression product and user based models average results are shown in Tables \ref{tab:logit_product_av_res} and \ref{tab:logit_user_av_res}.


\begin{table}[htp] \centering
\caption{SVM Product Based Models Average \(F_{1}\) scores.}\label{tab:svm_product_av_res}
\resizebox{0.9\textwidth}{!}{%
\begin{tabular}{|=l|+c|+r|+r|+r|+r|+r|+r|} \hline
\textbf{\begin{tabular}[c]{@{}l@{}}Prediction\\ Target\end{tabular}} 
    & \textbf{Category} 
    & \textbf{\begin{tabular}[c]{@{}l@{}}Advert \\ Viewing\\ Weekday \\ Time \\ Slots \end{tabular}} 
    & \textbf{\begin{tabular}[c]{@{}l@{}}Advert \\ Viewing\\ Weekday \\ Only\end{tabular}} 
    & \textbf{Demographics} 
    & \textbf{\begin{tabular}[c]{@{}l@{}}Advert \\ Viewing\\ Weekday \\ Time Slots and \\ Demographics\end{tabular}} 
    & \textbf{\begin{tabular}[c]{@{}l@{}}Advert \\ Viewing\\ Weekday \\ Only and \\ Demographics\end{tabular}} 
    & \textbf{\begin{tabular}[c]{@{}l@{}}Total \\ Average\end{tabular}} \\ \hline

\rowstyle{\bfseries}
\multirow{7}{*}{\textbf{\begin{tabular}[c]{@{}l@{}}Actual\\ Purchase\end{tabular}}}
    & \textbf{\begin{tabular}[c]{@{}l@{}}General\\ Average\end{tabular}} 
                & 0.293 & 0.292 & 0.489 & 0.495 & 0.497 & 0.413  \\ \cline{2-8} 
            & 0 & 0.000 & 0.000 & 0.211 & 0.225 & 0.218 & 0.131  \\ \cline{2-8} 
            & 1 & 0.856 & 0.852 & 0.876 & 0.878 & 0.875 & 0.867  \\ \cline{2-8} 
            & 2 & 0.000 & 0.000 & 0.327 & 0.343 & 0.343 & 0.204  \\ \cline{2-8} 
            & 3 & 0.000 & 0.000 & 0.161 & 0.171 & 0.171 & 0.102  \\ \cline{2-8} 
            & 4 & 0.000 & 0.000 & 0.446 & 0.441 & 0.441 & 0.267  \\ \cline{2-8} 
            & 5 & 0.901 & 0.900 & 0.910 & 0.911 & 0.911 & 0.907  \\ \hline

\rowstyle{\bfseries}
\multirow{7}{*}{\textbf{\begin{tabular}[c]{@{}l@{}}Purchase\\ Intention\end{tabular}}} 
    & \textbf{\begin{tabular}[c]{@{}l@{}}General\\ Average\end{tabular}} 
                & 0.252 & 0.248 & 0.273 & 0.275 & 0.276 & 0.265  \\ \cline{2-8} 
            & 0 & 0.000 & 0.000 & 0.000 & 0.000 & 0.000 & 0.000  \\ \cline{2-8} 
            & 1 & 0.570 & 0.558 & 0.590 & 0.590 & 0.596 & 0.581  \\ \cline{2-8} 
            & 2 & 0.000 & 0.000 & 0.000 & 0.000 & 0.000 & 0.000  \\ \cline{2-8} 
            & 3 & 0.115 & 0.110 & 0.139 & 0.139 & 0.138 & 0.129  \\ \cline{2-8} 
            & 4 & 0.166 & 0.156 & 0.226 & 0.226 & 0.233 & 0.202  \\ \cline{2-8} 
            & 5 & 0.662 & 0.666 & 0.685 & 0.685 & 0.689 & 0.678  \\ \hline
\rowstyle{\bfseries}
\textbf{\begin{tabular}[c]{@{}l@{}}Both \\ Targets\end{tabular}} & \textbf{\begin{tabular}[c]{@{}l@{}}Total\\ Average\end{tabular}}
                & 0.273 & 0.270 & 0.381 & 0.385 & 0.386 & 0.339  \\ \hline
\end{tabular}%
}
\end{table}

\begin{table}[htp] \centering
\caption{SVM User Based Models Average \(F_{1}\) scores.}\label{tab:svm_user_av_res}
\resizebox{0.9\textwidth}{!}{%
\begin{tabular}{|=l|+c|+r|+r|+r|+r|+r|+r|} \hline
\textbf{\begin{tabular}[c]{@{}l@{}}Prediction\\ Target\end{tabular}} 
    & \textbf{Category} 
    & \textbf{\begin{tabular}[c]{@{}l@{}}Advert \\ Viewing\\ Weekday \\ Time \\ Slots \end{tabular}} 
    & \textbf{\begin{tabular}[c]{@{}l@{}}Advert \\ Viewing\\ Weekday \\ Only\end{tabular}} 
    & \textbf{Demographics} 
    & \textbf{\begin{tabular}[c]{@{}l@{}}Advert \\ Viewing\\ Weekday \\ Time Slots and \\ Demographics\end{tabular}} 
    & \textbf{\begin{tabular}[c]{@{}l@{}}Advert \\ Viewing\\ Weekday \\ Only and \\ Demographics\end{tabular}} 
    & \textbf{\begin{tabular}[c]{@{}l@{}}Total \\ Average\end{tabular}} \\ \hline

\rowstyle{\bfseries}
\multirow{7}{*}{\textbf{\begin{tabular}[c]{@{}l@{}}Actual\\ Purchase\end{tabular}}}
    & \textbf{\begin{tabular}[c]{@{}l@{}}General\\ Average\end{tabular}} 
                & 0.317 & 0.315 & 0.391 & 0.359 & 0.373 & 0.351  \\ \cline{2-8} 
            & 0 & 0.055 & 0.049 & 0.095 & 0.085 & 0.087 & 0.074  \\ \cline{2-8} 
            & 1 & 0.745 & 0.755 & 0.854 & 0.789 & 0.812 & 0.791  \\ \cline{2-8} 
            & 2 & 0.074 & 0.070 & 0.140 & 0.114 & 0.126 & 0.105  \\ \cline{2-8} 
            & 3 & 0.076 & 0.071 & 0.108 & 0.107 & 0.121 & 0.096  \\ \cline{2-8} 
            & 4 & 0.140 & 0.126 & 0.254 & 0.221 & 0.239 & 0.196  \\ \cline{2-8} 
            & 5 & 0.812 & 0.822 & 0.893 & 0.840 & 0.855 & 0.844  \\ \hline

\rowstyle{\bfseries}
\multirow{7}{*}{\textbf{\begin{tabular}[c]{@{}l@{}}Purchase\\ Intention\end{tabular}}} 
    & \textbf{\begin{tabular}[c]{@{}l@{}}General\\ Average\end{tabular}} 
                & 0.297 & 0.289 & 0.242 & 0.296 & 0.290 & 0.283  \\ \cline{2-8} 
            & 0 & 0.036 & 0.032 & 0.006 & 0.036 & 0.033 & 0.029  \\ \cline{2-8} 
            & 1 & 0.553 & 0.553 & 0.534 & 0.548 & 0.553 & 0.548  \\ \cline{2-8} 
            & 2 & 0.048 & 0.043 & 0.007 & 0.049 & 0.042 & 0.038  \\ \cline{2-8} 
            & 3 & 0.217 & 0.195 & 0.093 & 0.217 & 0.198 & 0.184  \\ \cline{2-8} 
            & 4 & 0.301 & 0.279 & 0.174 & 0.299 & 0.280 & 0.267  \\ \cline{2-8} 
            & 5 & 0.629 & 0.634 & 0.639 & 0.626 & 0.632 & 0.632  \\ \hline
\rowstyle{\bfseries}
\textbf{\begin{tabular}[c]{@{}l@{}}Both \\ Targets\end{tabular}} & \textbf{\begin{tabular}[c]{@{}l@{}}Total\\ Average\end{tabular}}
                & 0.307 & 0.302 & 0.316 & 0.327 & 0.331 & 0.317  \\ \hline
\end{tabular}%
}
\end{table}


\begin{table}[htp] \centering
\caption{XGBoost Product Based Models Average \(F_{1}\) scores.}\label{tab:xgboost_product_av_res}
\resizebox{0.9\textwidth}{!}{%
\begin{tabular}{|=l|+c|+r|+r|+r|+r|+r|+r|} \hline
\textbf{\begin{tabular}[c]{@{}l@{}}Prediction\\ Target\end{tabular}} 
    & \textbf{Category} 
    & \textbf{\begin{tabular}[c]{@{}l@{}}Advert \\ Viewing\\ Weekday \\ Time \\ Slots \end{tabular}} 
    & \textbf{\begin{tabular}[c]{@{}l@{}}Advert \\ Viewing\\ Weekday \\ Only\end{tabular}} 
    & \textbf{Demographics} 
    & \textbf{\begin{tabular}[c]{@{}l@{}}Advert \\ Viewing\\ Weekday \\ Time Slots and \\ Demographics\end{tabular}} 
    & \textbf{\begin{tabular}[c]{@{}l@{}}Advert \\ Viewing\\ Weekday \\ Only and \\ Demographics\end{tabular}} 
    & \textbf{\begin{tabular}[c]{@{}l@{}}Total \\ Average\end{tabular}} \\ \hline

\rowstyle{\bfseries}
\multirow{7}{*}{\textbf{\begin{tabular}[c]{@{}l@{}}Actual\\ Purchase\end{tabular}}}
    & \textbf{\begin{tabular}[c]{@{}l@{}}General\\ Average\end{tabular}} 
                & 0.293 & 0.294 & 0.307 & 0.309 & 0.308 & 0.302  \\ \cline{2-8} 
            & 0 & 0.001 & 0.000 & 0.027 & 0.030 & 0.029 & 0.017  \\ \cline{2-8} 
            & 1 & 0.847 & 0.849 & 0.756 & 0.755 & 0.752 & 0.792  \\ \cline{2-8} 
            & 2 & 0.001 & 0.000 & 0.048 & 0.054 & 0.052 & 0.031  \\ \cline{2-8} 
            & 3 & 0.003 & 0.003 & 0.043 & 0.044 & 0.045 & 0.028  \\ \cline{2-8} 
            & 4 & 0.008 & 0.010 & 0.136 & 0.136 & 0.133 & 0.085  \\ \cline{2-8} 
            & 5 & 0.898 & 0.900 & 0.835 & 0.835 & 0.833 & 0.860  \\ \hline

\rowstyle{\bfseries}
\multirow{7}{*}{\textbf{\begin{tabular}[c]{@{}l@{}}Purchase\\ Intention\end{tabular}}} 
    & \textbf{\begin{tabular}[c]{@{}l@{}}General\\ Average\end{tabular}} 
                & 0.257 & 0.257 & 0.266 & 0.266 & 0.268 & 0.263  \\ \cline{2-8} 
            & 0 & 0.000 & 0.000 & 0.000 & 0.002 & 0.001 & 0.001  \\ \cline{2-8} 
            & 1 & 0.574 & 0.571 & 0.573 & 0.565 & 0.572 & 0.571  \\ \cline{2-8} 
            & 2 & 0.000 & 0.001 & 0.001 & 0.001 & 0.002 & 0.001  \\ \cline{2-8} 
            & 3 & 0.121 & 0.122 & 0.136 & 0.136 & 0.142 & 0.131  \\ \cline{2-8} 
            & 4 & 0.175 & 0.175 & 0.211 & 0.220 & 0.219 & 0.200  \\ \cline{2-8} 
            & 5 & 0.670 & 0.674 & 0.673 & 0.669 & 0.674 & 0.672  \\ \hline
\rowstyle{\bfseries}
\textbf{\begin{tabular}[c]{@{}l@{}}Both \\ Targets\end{tabular}} & \textbf{\begin{tabular}[c]{@{}l@{}}Total\\ Average\end{tabular}}
                & 0.275 & 0.275 & 0.287 & 0.287 & 0.288 & 0.282  \\ \hline
\end{tabular}%
}
\end{table}

\begin{table}[htp] \centering
\caption{XGBoost User Based Models Average \(F_{1}\) scores.}\label{tab:xgboost_user_av_res}
\resizebox{0.9\textwidth}{!}{%
\begin{tabular}{|=l|+c|+r|+r|+r|+r|+r|+r|} \hline
\textbf{\begin{tabular}[c]{@{}l@{}}Prediction\\ Target\end{tabular}} 
    & \textbf{Category} 
    & \textbf{\begin{tabular}[c]{@{}l@{}}Advert \\ Viewing\\ Weekday \\ Time \\ Slots \end{tabular}} 
    & \textbf{\begin{tabular}[c]{@{}l@{}}Advert \\ Viewing\\ Weekday \\ Only\end{tabular}} 
    & \textbf{Demographics} 
    & \textbf{\begin{tabular}[c]{@{}l@{}}Advert \\ Viewing\\ Weekday \\ Time Slots and \\ Demographics\end{tabular}} 
    & \textbf{\begin{tabular}[c]{@{}l@{}}Advert \\ Viewing\\ Weekday \\ Only and \\ Demographics\end{tabular}} 
    & \textbf{\begin{tabular}[c]{@{}l@{}}Total \\ Average\end{tabular}} \\ \hline

\rowstyle{\bfseries}
\multirow{7}{*}{\textbf{\begin{tabular}[c]{@{}l@{}}Actual\\ Purchase\end{tabular}}}
    & \textbf{\begin{tabular}[c]{@{}l@{}}General\\ Average\end{tabular}} 
                & 0.291 & 0.291 & 0.297 & 0.301 & 0.302 & 0.296  \\ \cline{2-8} 
            & 0 & 0.008 & 0.008 & 0.018 & 0.021 & 0.021 & 0.015  \\ \cline{2-8} 
            & 1 & 0.771 & 0.769 & 0.752 & 0.758 & 0.760 & 0.762  \\ \cline{2-8} 
            & 2 & 0.014 & 0.014 & 0.030 & 0.032 & 0.033 & 0.025  \\ \cline{2-8} 
            & 3 & 0.027 & 0.029 & 0.042 & 0.048 & 0.049 & 0.039  \\ \cline{2-8} 
            & 4 & 0.082 & 0.084 & 0.119 & 0.121 & 0.123 & 0.106  \\ \cline{2-8} 
            & 5 & 0.841 & 0.842 & 0.825 & 0.827 & 0.825 & 0.832  \\ \hline

\rowstyle{\bfseries}
\multirow{7}{*}{\textbf{\begin{tabular}[c]{@{}l@{}}Purchase\\ Intention\end{tabular}}} 
    & \textbf{\begin{tabular}[c]{@{}l@{}}General\\ Average\end{tabular}} 
                & 0.267 & 0.269 & 0.239 & 0.267 & 0.268 & 0.262  \\ \cline{2-8} 
            & 0 & 0.008 & 0.008 & 0.001 & 0.008 & 0.007 & 0.006  \\ \cline{2-8} 
            & 1 & 0.538 & 0.543 & 0.533 & 0.542 & 0.542 & 0.540  \\ \cline{2-8} 
            & 2 & 0.013 & 0.013 & 0.001 & 0.013 & 0.013 & 0.010  \\ \cline{2-8} 
            & 3 & 0.154 & 0.161 & 0.085 & 0.155 & 0.159 & 0.143  \\ \cline{2-8} 
            & 4 & 0.249 & 0.250 & 0.174 & 0.248 & 0.252 & 0.235  \\ \cline{2-8} 
            & 5 & 0.638 & 0.637 & 0.643 & 0.636 & 0.636 & 0.638  \\ \hline
                
\rowstyle{\bfseries}
\textbf{\begin{tabular}[c]{@{}l@{}}Both \\ Targets\end{tabular}} & \textbf{\begin{tabular}[c]{@{}l@{}}Total\\ Average\end{tabular}}
                & 0.279 & 0.280 & 0.268 & 0.284 & 0.285 & 0.279  \\ \hline
\end{tabular}%
}
\end{table}


\begin{table}[htp] \centering
\caption{Logistic Regression Product Based Models Average \(F_{1}\) scores.}\label{tab:logit_product_av_res}
\resizebox{0.9\textwidth}{!}{%
\begin{tabular}{|=l|+c|+r|+r|+r|+r|+r|+r|} \hline
\textbf{\begin{tabular}[c]{@{}l@{}}Prediction\\ Target\end{tabular}} 
    & \textbf{Category} 
    & \textbf{\begin{tabular}[c]{@{}l@{}}Advert \\ Viewing\\ Weekday \\ Time \\ Slots \end{tabular}} 
    & \textbf{\begin{tabular}[c]{@{}l@{}}Advert \\ Viewing\\ Weekday \\ Only\end{tabular}} 
    & \textbf{Demographics} 
    & \textbf{\begin{tabular}[c]{@{}l@{}}Advert \\ Viewing\\ Weekday \\ Time Slots and \\ Demographics\end{tabular}} 
    & \textbf{\begin{tabular}[c]{@{}l@{}}Advert \\ Viewing\\ Weekday \\ Only and \\ Demographics\end{tabular}} 
    & \textbf{\begin{tabular}[c]{@{}l@{}}Total \\ Average\end{tabular}} \\ \hline

\rowstyle{\bfseries}
\multirow{7}{*}{\textbf{\begin{tabular}[c]{@{}l@{}}Actual\\ Purchase\end{tabular}}}
    & \textbf{\begin{tabular}[c]{@{}l@{}}General\\ Average\end{tabular}} 
                & 0.293 & 0.293 & 0.500 & 0.513 & 0.508 & 0.421  \\ \cline{2-8} 
            & 0 & 0.000 & 0.000 & 0.226 & 0.234 & 0.237 & 0.139  \\ \cline{2-8} 
            & 1 & 0.851 & 0.853 & 0.874 & 0.874 & 0.875 & 0.865  \\ \cline{2-8} 
            & 2 & 0.000 & 0.000 & 0.343 & 0.368 & 0.355 & 0.213  \\ \cline{2-8} 
            & 3 & 0.002 & 0.000 & 0.181 & 0.208 & 0.195 & 0.117  \\ \cline{2-8} 
            & 4 & 0.006 & 0.004 & 0.462 & 0.480 & 0.475 & 0.286  \\ \cline{2-8} 
            & 5 & 0.899 & 0.901 & 0.914 & 0.914 & 0.914 & 0.909  \\ \hline

\rowstyle{\bfseries}
\multirow{7}{*}{\textbf{\begin{tabular}[c]{@{}l@{}}Purchase\\ Intention\end{tabular}}} 
    & \textbf{\begin{tabular}[c]{@{}l@{}}General\\ Average\end{tabular}} 
                & 0.259 & 0.256 & 0.289 & 0.294 & 0.292 & 0.278  \\ \cline{2-8} 
            & 0 & 0.000 & 0.000 & 0.000 & 0.000 & 0.000 & 0.000  \\ \cline{2-8} 
            & 1 & 0.575 & 0.572 & 0.608 & 0.611 & 0.611 & 0.595  \\ \cline{2-8} 
            & 2 & 0.000 & 0.000 & 0.000 & 0.001 & 0.000 & 0.000  \\ \cline{2-8} 
            & 3 & 0.125 & 0.118 & 0.159 & 0.170 & 0.163 & 0.147  \\ \cline{2-8} 
            & 4 & 0.184 & 0.174 & 0.270 & 0.280 & 0.278 & 0.237  \\ \cline{2-8} 
            & 5 & 0.671 & 0.668 & 0.696 & 0.703 & 0.701 & 0.688  \\ \hline
\rowstyle{\bfseries}
\textbf{\begin{tabular}[c]{@{}l@{}}Both \\ Targets\end{tabular}} & \textbf{\begin{tabular}[c]{@{}l@{}}Total\\ Average\end{tabular}}
                & 0.276 & 0.274 & 0.394 & 0.404 & 0.400 & 0.350  \\ \hline
\end{tabular}%
}
\end{table}

\begin{table}[htp] \centering
\caption{Logistic Regression User Based Models Average \(F_{1}\) scores.}\label{tab:logit_user_av_res}
\resizebox{0.9\textwidth}{!}{%
\begin{tabular}{|=l|+c|+r|+r|+r|+r|+r|+r|} \hline
\textbf{\begin{tabular}[c]{@{}l@{}}Prediction\\ Target\end{tabular}} 
    & \textbf{Category} 
    & \textbf{\begin{tabular}[c]{@{}l@{}}Advert \\ Viewing\\ Weekday \\ Time \\ Slots \end{tabular}} 
    & \textbf{\begin{tabular}[c]{@{}l@{}}Advert \\ Viewing\\ Weekday \\ Only\end{tabular}} 
    & \textbf{Demographics} 
    & \textbf{\begin{tabular}[c]{@{}l@{}}Advert \\ Viewing\\ Weekday \\ Time Slots and \\ Demographics\end{tabular}} 
    & \textbf{\begin{tabular}[c]{@{}l@{}}Advert \\ Viewing\\ Weekday \\ Only and \\ Demographics\end{tabular}} 
    & \textbf{\begin{tabular}[c]{@{}l@{}}Total \\ Average\end{tabular}} \\ \hline

\rowstyle{\bfseries}
\multirow{7}{*}{\textbf{\begin{tabular}[c]{@{}l@{}}Actual\\ Purchase\end{tabular}}}
    & \textbf{\begin{tabular}[c]{@{}l@{}}General\\ Average\end{tabular}} 
                & 0.312 & 0.309 & 0.347 & 0.333 & 0.341 & 0.328  \\ \cline{2-8} 
            & 0 & 0.046 & 0.040 & 0.037 & 0.056 & 0.054 & 0.047  \\ \cline{2-8} 
            & 1 & 0.740 & 0.750 & 0.846 & 0.771 & 0.794 & 0.780  \\ \cline{2-8} 
            & 2 & 0.064 & 0.058 & 0.070 & 0.083 & 0.082 & 0.071  \\ \cline{2-8} 
            & 3 & 0.074 & 0.066 & 0.064 & 0.089 & 0.091 & 0.077  \\ \cline{2-8} 
            & 4 & 0.139 & 0.124 & 0.175 & 0.173 & 0.185 & 0.159  \\ \cline{2-8} 
            & 5 & 0.806 & 0.815 & 0.888 & 0.827 & 0.844 & 0.836  \\ \hline

\rowstyle{\bfseries}
\multirow{7}{*}{\textbf{\begin{tabular}[c]{@{}l@{}}Purchase\\ Intention\end{tabular}}} 
    & \textbf{\begin{tabular}[c]{@{}l@{}}General\\ Average\end{tabular}} 
                & 0.298 & 0.295 & 0.242 & 0.298 & 0.293 & 0.285  \\ \cline{2-8} 
            & 0 & 0.037 & 0.035 & 0.007 & 0.038 & 0.034 & 0.030  \\ \cline{2-8} 
            & 1 & 0.555 & 0.557 & 0.535 & 0.554 & 0.556 & 0.551  \\ \cline{2-8} 
            & 2 & 0.052 & 0.047 & 0.009 & 0.052 & 0.044 & 0.041  \\ \cline{2-8} 
            & 3 & 0.220 & 0.204 & 0.093 & 0.220 & 0.207 & 0.189  \\ \cline{2-8} 
            & 4 & 0.298 & 0.288 & 0.175 & 0.300 & 0.286 & 0.269  \\ \cline{2-8} 
            & 5 & 0.628 & 0.635 & 0.636 & 0.628 & 0.634 & 0.632  \\ \hline
\rowstyle{\bfseries}
\textbf{\begin{tabular}[c]{@{}l@{}}Both \\ Targets\end{tabular}} & \textbf{\begin{tabular}[c]{@{}l@{}}Total\\ Average\end{tabular}}
                & 0.305 & 0.302 & 0.294 & 0.316 & 0.317 & 0.307  \\ \hline
\end{tabular}%
}
\end{table}

\subsection{Statistical analysis}
\label{h_res}

In this study we performed a series of experiments where we trained different prediction models based on either advert viewing time or demographic data to predict purchase behaviors of actual purchase and purchase intention across 3000 users and 36 products. If we were able to predict purchase behaviors with models based on exposure time more reliably than with models based on demographic data, the obvious strategy for businesses would be to increase the number of adverts. On the other hand, if models based on exposure time had unreliable predictability in contrast to models based on demographic data, doubts would surface about the effectiveness of the hard investment in television advertising. 

In order to analyze the change in predictability of purchase behavior we averaged the results of predictions across different variations of the experiments, detailed in the previous section, and then performed \(t\)-tests to observe the difference in performance between sets of results. We established 3 hypotheses to test for, explained below.

\begin{hyp}
\label{hyp:1}
Advert viewing time based models perform differently from demographics based models.
\end{hyp}

For this hypothesis, we performed a \(t\)-test using the results from models that include advert viewing time and the models that only include demographic data. More specifically, we tested the Weekday Time Slot model results against the Demographics models, and the Weekday Only models against the Demographics models. The p-values for each \(t\)-test are shown in Table \ref{tab:h1_ap} for Actual Purchase predictions and in Table \ref{tab:h1_pi} for Purchase Intention predictions. With these tests, we will examine the changes in predictability against demographic data, which we are using as the control data for our experiments. This will allow us to determine whether the advert viewing time based models are performing better or worse than the demographic models, and therefore conclude whether the advert viewing time is having an effect on customers purchase behavior or if it is decided by external factors.

\begin{table}[p] \centering
\caption{Hypothesis 1 \(t\)-test: p-values for Actual Purchase behavior.}\label{tab:h1_ap}
\resizebox{1\textwidth}{!}{%
\begin{tabular}{|c|c|l|r|r|r|r|r|r|}
\hline
\multirow{2}{*}{\textbf{Model}} 
    & \multirow{2}{*}{\textbf{Base}} 
    & \multicolumn{1}{c|}{\multirow{2}{*}{\textbf{Configuration}}} 
    & \multicolumn{6}{c|}{\textbf{Actual Purchase Categories}} \\ \cline{4-9} 
    & & \multicolumn{1}{c|}{}
      & \multicolumn{1}{c|}{\textbf{0}} 
      & \multicolumn{1}{c|}{\textbf{1}} 
      & \multicolumn{1}{c|}{\textbf{2}} 
      & \multicolumn{1}{c|}{\textbf{3}} 
      & \multicolumn{1}{c|}{\textbf{4}} 
      & \multicolumn{1}{c|}{\textbf{5}} \\ \hline
\multirow{4}{*}{SVM} 
    & \multirow{2}{*}{product} 
        & Weekday Time Slot     & 0.000 & 0.328 & 0.000 & 0.001 & 0.000 & 0.567 \\ \cline{3-9} 
      & & Weekday Only              & 0.000 & 0.284 & 0.000 & 0.001 & 0.000 & 0.527 \\ \cline{2-9} 
    & \multirow{2}{*}{user} 
        & Weekday Time Slot     & 0.000 & 0.000 & 0.000 & 0.000 & 0.000 & 0.000 \\ \cline{3-9} 
      & & Weekday Only              & 0.000 & 0.000 & 0.000 & 0.000 & 0.000 & 0.000 \\ \hline
\multirow{4}{*}{XGBoost}
    & \multirow{2}{*}{product} 
        & Weekday Time Slot     & 0.000 & 0.005 & 0.000 & 0.015 & 0.000 & 0.013 \\ \cline{3-9} 
      & & Weekday Only              & 0.000 & 0.004 & 0.000 & 0.015 & 0.000 & 0.011 \\ \cline{2-9} 
    & \multirow{2}{*}{user}    
        & Weekday Time Slot     & 0.000 & 0.004 & 0.000 & 0.000 & 0.000 & 0.004 \\ \cline{3-9} 
      & & Weekday Only              & 0.000 & 0.010 & 0.000 & 0.000 & 0.000 & 0.003 \\ \hline
\multirow{4}{*}{\begin{tabular}[c]{@{}c@{}}Logistic \\ Regression\end{tabular}} 
    & \multirow{2}{*}{product} 
        & Weekday Time Slot     & 0.000 & 0.282 & 0.000 & 0.000 & 0.000 & 0.348 \\ \cline{3-9} 
      & & Weekday Only              & 0.000 & 0.327 & 0.000 & 0.000 & 0.000 & 0.414 \\ \cline{2-9} 
    & \multirow{2}{*}{user} 
        & Weekday Time Slot     & 0.012 & 0.000 & 0.236 & 0.032 & 0.000 & 0.000 \\ \cline{3-9} 
      & & Weekday Only              & 0.374 & 0.000 & 0.007 & 0.757 & 0.000 & 0.000 \\ \hline
\end{tabular}%
}
\end{table}

\begin{table}[p] \centering
\caption{Hypothesis 1 \(t\)-test: p-values for Purchase Intention behavior.}\label{tab:h1_pi}
\resizebox{1\textwidth}{!}{%
\begin{tabular}{|c|c|l|r|r|r|r|r|r|}
\hline
\multirow{2}{*}{\textbf{Model}} 
    & \multirow{2}{*}{\textbf{Base}} 
    & \multicolumn{1}{c|}{\multirow{2}{*}{\textbf{Configuration}}} 
    & \multicolumn{6}{c|}{\textbf{Purchase Intention Categories}} \\ \cline{4-9} 
    & & \multicolumn{1}{c|}{}
      & \multicolumn{1}{c|}{\textbf{0}} 
      & \multicolumn{1}{c|}{\textbf{1}} 
      & \multicolumn{1}{c|}{\textbf{2}} 
      & \multicolumn{1}{c|}{\textbf{3}} 
      & \multicolumn{1}{c|}{\textbf{4}} 
      & \multicolumn{1}{c|}{\textbf{5}} \\ \hline
\multirow{4}{*}{SVM}
    & \multirow{2}{*}{product} 
        & Weekday Time Slot     & nan   & 0.803 & nan   & 0.700 & 0.405 & 0.767 \\ \cline{3-9} 
      & & Weekday Only              & nan   & 0.694 & nan   & 0.644 & 0.329 & 0.808 \\ \cline{2-9} 
    & \multirow{2}{*}{user}
        & Weekday Time Slot     & 0.000 & 0.026 & 0.000 & 0.000 & 0.000 & 0.234 \\ \cline{3-9} 
      & & Weekday Only              & 0.000 & 0.028 & 0.000 & 0.000 & 0.000 & 0.534 \\ \hline
\multirow{4}{*}{XGBoost}
    & \multirow{2}{*}{product}
        & Weekday Time Slot     & 0.324 & 0.983 & 0.041 & 0.804 & 0.598 & 0.969 \\ \cline{3-9} 
      & & Weekday Only              & 0.324 & 0.982 & 0.203 & 0.811 & 0.606 & 0.992 \\ \cline{2-9} 
    & \multirow{2}{*}{user}
        & Weekday Time Slot     & 0.000 & 0.533 & 0.000 & 0.000 & 0.000 & 0.585 \\ \cline{3-9} 
      & & Weekday Only              & 0.000 & 0.246 & 0.000 & 0.000 & 0.000 & 0.509 \\ \hline
\multirow{4}{*}{\begin{tabular}[c]{@{}c@{}}Logistic \\ Regression\end{tabular}}
    & \multirow{2}{*}{product}
        & Weekday Time Slot     & nan   & 0.673 & 0.803 & 0.582 & 0.222 & 0.724 \\ \cline{3-9} 
      & & Weekday Only              & nan   & 0.655 & 0.324 & 0.514 & 0.175 & 0.704 \\ \cline{2-9} 
    & \multirow{2}{*}{user}
        & Weekday Time Slot     & 0.000 & 0.017 & 0.000 & 0.000 & 0.000 & 0.364 \\ \cline{3-9} 
      & & Weekday Only              & 0.000 & 0.010 & 0.000 & 0.000 & 0.000 & 0.947 \\ \hline
\end{tabular}%
}
\end{table}

\begin{hyp}
\label{hyp:2}
Demographic and advert viewing based models perform differently from demographic based models.
\end{hyp}

For this hypothesis, we performed a \(t\)-test bewteen the results from models that include both advert viewing time and demographic data, and the models that only include demographic data. More specifically, we tested the Weekday Time Slot and Demographics model results against the Demographics models, and the Weekday and Demographics models against the Demographics models. The p-values for each \(t\)-test are shown in Table \ref{tab:h2_ap} for Actual Purchase predictions and in Table \ref{tab:h2_pi} for Purchase Intention predictions. With these tests, we will examine if adding the advert viewing data to the demographic data causes any major changes, to determine if the predictions are being improved, worsened, or if they stay the same regardless of advert viewing.

\begin{table}[p] \centering
\caption{Hypothesis 2 \(t\)-test: p-values for Actual Purchase behavior.}\label{tab:h2_ap}
\resizebox{\textwidth}{!}{%
\begin{tabular}{|c|c|l|r|r|r|r|r|r|}
\hline
\multirow{2}{*}{\textbf{Model}} 
    & \multirow{2}{*}{\textbf{Base}} 
    & \multicolumn{1}{c|}{\multirow{2}{*}{\textbf{Configuration}}} 
    & \multicolumn{6}{c|}{\textbf{Actual Purchase Categories}} \\ \cline{4-9} 
    & & \multicolumn{1}{c|}{}
      & \multicolumn{1}{c|}{\textbf{0}} 
      & \multicolumn{1}{c|}{\textbf{1}} 
      & \multicolumn{1}{c|}{\textbf{2}} 
      & \multicolumn{1}{c|}{\textbf{3}} 
      & \multicolumn{1}{c|}{\textbf{4}} 
      & \multicolumn{1}{c|}{\textbf{5}} \\ \hline
\multirow{4}{*}{SVM} 
    & \multirow{2}{*}{product} 
        & Weekday Time Slot     & 0.923 & 0.953 & 0.748 & 0.777 & 0.968 & 0.980 \\ \cline{3-9} 
      & & Weekday Only              & 0.850 & 0.936 & 0.839 & 0.868 & 0.942 & 0.956 \\ \cline{2-9} 
 & \multirow{2}{*}{user} 
    & Weekday Time Slot         & 0.056 & 0.000 & 0.000 & 0.837 & 0.000 & 0.000 \\ \cline{3-9} 
      & & Weekday Only              & 0.139 & 0.000 & 0.033 & 0.038 & 0.078 & 0.000 \\ \hline
\multirow{4}{*}{XGBoost} 
    & \multirow{2}{*}{product} 
        & Weekday Time Slot     & 0.758 & 0.973 & 0.581 & 0.936 & 0.991 & 0.993 \\ \cline{3-9} 
      & & Weekday Only              & 0.811 & 0.929 & 0.703 & 0.897 & 0.924 & 0.956 \\ \cline{2-9} 
 & \multirow{2}{*}{user} 
    & Weekday Time Slot         & 0.074 & 0.318 & 0.413 & 0.057 & 0.625 & 0.655 \\ \cline{3-9} 
      & & Weekday Only              & 0.069 & 0.187 & 0.173 & 0.045 & 0.388 & 0.993 \\ \hline
\multirow{4}{*}{\begin{tabular}[c]{@{}c@{}}Logistic \\ Regression\end{tabular}} 
    & \multirow{2}{*}{product} 
        & Weekday Time Slot     & 0.894 & 0.997 & 0.674 & 0.627 & 0.772 & 0.971 \\ \cline{3-9} 
      & & Weekday Only              & 0.856 & 0.955 & 0.846 & 0.808 & 0.831 & 0.990 \\ \cline{2-9} 
 & \multirow{2}{*}{user} 
    & Weekday Time Slot         & 0.000 & 0.000 & 0.007 & 0.000 & 0.823 & 0.000 \\ \cline{3-9} 
      & & Weekday Only              & 0.000 & 0.000 & 0.017 & 0.000 & 0.187 & 0.000 \\ \hline
\end{tabular}%
}
\end{table}

\begin{table}[p] \centering
\caption{Hypothesis 2 \(t\)-test: p-values for Purchase Intention behavior.}\label{tab:h2_pi}
\resizebox{\textwidth}{!}{%
\begin{tabular}{|c|c|l|r|r|r|r|r|r|}
\hline
\multirow{2}{*}{\textbf{Model}} 
    & \multirow{2}{*}{\textbf{Base}} 
    & \multicolumn{1}{c|}{\multirow{2}{*}{\textbf{Configuration}}} 
    & \multicolumn{6}{c|}{\textbf{Purchase Intention Categories}} \\ \cline{4-9} 
    & & \multicolumn{1}{c|}{}
      & \multicolumn{1}{c|}{\textbf{0}} 
      & \multicolumn{1}{c|}{\textbf{1}} 
      & \multicolumn{1}{c|}{\textbf{2}} 
      & \multicolumn{1}{c|}{\textbf{3}} 
      & \multicolumn{1}{c|}{\textbf{4}} 
      & \multicolumn{1}{c|}{\textbf{5}} \\ \hline
\multirow{4}{*}{SVM} 
    & \multirow{2}{*}{product} 
        & Weekday Time Slot      & nan & 0.942 & nan & 0.985 & 0.930 & 0.969 \\ \cline{3-9} 
      & & Weekday Only               & nan & 0.986 & nan & 0.991 & 0.950 & 0.955 \\ \cline{2-9} 
    & \multirow{2}{*}{user} 
        & Weekday Time Slot      & 0.000 & 0.099 & 0.000 & 0.000 & 0.000 & 0.118 \\ \cline{3-9} 
      & & Weekday Only               & 0.000 & 0.029 & 0.000 & 0.000 & 0.000 & 0.402 \\ \hline
\multirow{4}{*}{XGBoost} 
    & \multirow{2}{*}{product} 
        & Weekday Time Slot      & 0.002 & 0.924 & 0.850 & 0.996 & 0.900 & 0.960 \\ \cline{3-9} 
      & & Weekday Only               & 0.115 & 0.990 & 0.364 & 0.916 & 0.916 & 0.986 \\ \cline{2-9} 
    & \multirow{2}{*}{user} 
        & Weekday Time Slot      & 0.000 & 0.290 & 0.000 & 0.000 & 0.000 & 0.409 \\ \cline{3-9} 
      & & Weekday Only               & 0.000 & 0.303 & 0.000 & 0.000 & 0.000 & 0.400 \\ \hline
\multirow{4}{*}{\begin{tabular}[c]{@{}c@{}}Logistic \\ Regression\end{tabular}} 
    & \multirow{2}{*}{product} 
        & Weekday Time Slot      & nan & 0.957 & 0.331 & 0.856 & 0.884 & 0.916 \\ \cline{3-9} 
      & & Weekday Only               & nan & 0.957 & 0.871 & 0.943 & 0.903 & 0.943 \\ \cline{2-9} 
    & \multirow{2}{*}{user} 
        & Weekday Time Slot      & 0.000 & 0.028 & 0.000 & 0.000 & 0.000 & 0.313 \\ \cline{3-9} 
      & & Weekday Only               & 0.000 & 0.016 & 0.000 & 0.000 & 0.000 & 0.869 \\ \hline
\end{tabular}%
}
\end{table}

\begin{hyp}
\label{hyp:3}
Advert viewing time based models perform differently from demographic and advert viewing based models.
\end{hyp}

For this hypothesis, we performed a \(t\)-test bewteen the results from models that include both advert viewing time and demographic data, and the models that only include advert viewing data. More specifically, we tested the Weekday Time Slot and Demographics model results against the Weekday Time Slot models, and the Weekday and Demographics models against the Weekday Only models. The p-values for each \(t\)-test are shown in Table \ref{tab:h3_ap} for Actual Purchase predictions and in Table \ref{tab:h3_pi} for Purchase Intention predictions. With these tests, we will examine if adding the demographic data to the advert viewing data causes any major changes, to determine if the predictions are being improved, worsened or if they stay the same regardless of demographic variances. By performing this last test, as well as the differences tested by Hypothesis \ref{hyp:1} and Hypothesis \ref{hyp:2}, we can assume significant differences across all major 3 groups of data.

\begin{table}[p] \centering
\caption{Hypothesis 3 \(t\)-test: p-values for Actual Purchase behavior.}\label{tab:h3_ap}
\resizebox{\textwidth}{!}{%
\begin{tabular}{|c|c|l|r|r|r|r|r|r|}
\hline
\multirow{2}{*}{\textbf{Model}} 
    & \multirow{2}{*}{\textbf{Base}} 
    & \multicolumn{1}{c|}{\multirow{2}{*}{\textbf{Configuration}}} 
    & \multicolumn{6}{c|}{\textbf{Actual Purchase Categories}} \\ \cline{4-9} 
    & & \multicolumn{1}{c|}{}
      & \multicolumn{1}{c|}{\textbf{0}} 
      & \multicolumn{1}{c|}{\textbf{1}} 
      & \multicolumn{1}{c|}{\textbf{2}} 
      & \multicolumn{1}{c|}{\textbf{3}} 
      & \multicolumn{1}{c|}{\textbf{4}} 
      & \multicolumn{1}{c|}{\textbf{5}} \\ \hline
\multirow{4}{*}{SVM} 
    & \multirow{2}{*}{product}
        & Weekday Time Slot     & 0.000 & 0.354 & 0.000 & 0.000 & 0.000 & 0.553 \\ \cline{3-9} 
      & & Weekday Only              & 0.000 & 0.253 & 0.000 & 0.000 & 0.000 & 0.488 \\ \cline{2-9} 
    & \multirow{2}{*}{user}
        & Weekday Time Slot     & 0.000 & 0.000 & 0.000 & 0.000 & 0.000 & 0.000 \\ \cline{3-9} 
      & & Weekday Only              & 0.000 & 0.000 & 0.000 & 0.000 & 0.000 & 0.000 \\ \hline
\multirow{4}{*}{XGBoost} 
    & \multirow{2}{*}{product}
        & Weekday Time Slot     & 0.000 & 0.005 & 0.000 & 0.009 & 0.000 & 0.013 \\ \cline{3-9} 
      & & Weekday Only              & 0.000 & 0.003 & 0.000 & 0.013 & 0.000 & 0.009 \\ \cline{2-9} 
    & \multirow{2}{*}{user}
        & Weekday Time Slot     & 0.000 & 0.051 & 0.000 & 0.000 & 0.000 & 0.011 \\ \cline{3-9} 
      & & Weekday Only              & 0.000 & 0.181 & 0.000 & 0.000 & 0.000 & 0.002 \\ \hline
\multirow{4}{*}{\begin{tabular}[c]{@{}l@{}}Logistic\\ Regression\end{tabular}} 
    & \multirow{2}{*}{product}
        & Weekday Time Slot     & 0.000 & 0.287 & 0.000 & 0.000 & 0.000 & 0.367 \\ \cline{3-9} 
      & & Weekday Only              & 0.000 & 0.303 & 0.000 & 0.000 & 0.000 & 0.400 \\ \cline{2-9} 
    & \multirow{2}{*}{user}
        & Weekday Time Slot     & 0.005 & 0.000 & 0.000 & 0.001 & 0.000 & 0.000 \\ \cline{3-9} 
      & & Weekday Only              & 0.000 & 0.000 & 0.000 & 0.000 & 0.000 & 0.000 \\ \hline
\end{tabular}%
}
\end{table}

\begin{table}[p] \centering
\caption{Hypothesis 3 \(t\)-test: p-values for Purchase Intention behavior.}\label{tab:h3_pi}
\resizebox{\textwidth}{!}{%
\begin{tabular}{|c|c|l|r|r|r|r|r|r|}
\hline
\multirow{2}{*}{\textbf{Model}} 
    & \multirow{2}{*}{\textbf{Base}} 
    & \multicolumn{1}{c|}{\multirow{2}{*}{\textbf{Configuration}}} 
    & \multicolumn{6}{c|}{\textbf{Purchase Intention Categories}} \\ \cline{4-9} 
    & & \multicolumn{1}{c|}{}
      & \multicolumn{1}{c|}{\textbf{0}} 
      & \multicolumn{1}{c|}{\textbf{1}} 
      & \multicolumn{1}{c|}{\textbf{2}} 
      & \multicolumn{1}{c|}{\textbf{3}} 
      & \multicolumn{1}{c|}{\textbf{4}} 
      & \multicolumn{1}{c|}{\textbf{5}} \\ \hline
\multirow{4}{*}{SVM} 
    & \multirow{2}{*}{product} 
        & Weekday Time Slot     & nan & 0.750 & nan & 0.713 & 0.359 & 0.738 \\ \cline{3-9} 
      & & Weekday Only              & nan & 0.682 & nan & 0.633 & 0.300 & 0.766 \\ \cline{2-9} 
    & \multirow{2}{*}{user} 
        & Weekday Time Slot     & 0.874 & 0.494 & 0.853 & 0.934 & 0.835 & 0.655 \\ \cline{3-9} 
      & & Weekday Only              & 0.822 & 0.991 & 0.621 & 0.658 & 0.806 & 0.805 \\ \hline
\multirow{4}{*}{XGBoost} 
    & \multirow{2}{*}{product} 
        & Weekday Time Slot     & 0.031 & 0.910 & 0.071 & 0.805 & 0.514 & 0.993 \\ \cline{3-9} 
      & & Weekday Only              & 0.197 & 0.992 & 0.051 & 0.731 & 0.532 & 0.994 \\ \cline{2-9} 
    & \multirow{2}{*}{user} 
        & Weekday Time Slot     & 0.946 & 0.634 & 0.803 & 0.838 & 0.885 & 0.757 \\ \cline{3-9} 
      & & Weekday Only              & 0.586 & 0.886 & 0.941 & 0.764 & 0.880 & 0.841 \\ \hline
\multirow{4}{*}{\begin{tabular}[c]{@{}l@{}}Logistic\\ Regression\end{tabular}} 
    & \multirow{2}{*}{product} 
        & Weekday Time Slot     & nan & 0.634 & 0.427 & 0.464 & 0.169 & 0.647 \\ \cline{3-9} 
      & & Weekday Only              & nan & 0.618 & 0.324 & 0.468 & 0.139 & 0.652 \\ \cline{2-9} 
    & \multirow{2}{*}{user} 
        & Weekday Time Slot     & 0.904 & 0.823 & 0.934 & 0.968 & 0.746 & 0.905 \\ \cline{3-9} 
      & & Weekday Only              & 0.556 & 0.848 & 0.315 & 0.708 & 0.743 & 0.907 \\ \hline
\end{tabular}%
}
\end{table}

\section{Discussion}
\label{discussion}

\subsection{Influence of TV adverts on Actual Purchase and Purchase Intention}
\label{disc_ap_pi}

Observing our results across models in the Tables of section \ref{av_res}, in general, we can observe that SVM models perform relatively better than XGBoost models and Logistic Models, and that the differences and directional change in averages between Advert Viewing Time based models, Demographics based models, and Advert Viewing Time and Demographics based models stay consistent across SVM, XGBoost and Logistic Regression models. That is to say, low predictability in Advert Viewing Time based models compared to Demographic data models stays constant regardless of the changes in performance across prediction techniques.
  
In Tables \ref{tab:svm_product_av_res} and \ref{tab:svm_user_av_res} we can observe this more closely. In general for Actual Purchase behavior, predictions using Advert Viewing Time only have a lower performance than the other models. Specially in categories 2 and 4 of the purchase behavior, we can see that the average predictability rises from 0 or close to 0, to a higher predictability in every case that demographic data is used for positive purchase behavior.

We can confirm this increase is statistically significant by observing the results of Hypothesis \ref{hyp:1} in Table \ref{tab:h1_ap}. For the most part, excluding negative purchase behavior, the data is significantly different at a 95\% confidence level (\(p<0.05\)) between models that use advert viewing time as a base for prediction and models that use demographic data as a base for prediction. Moreover, we can confirm that the changes in predictability between models that include both advert viewing time and demographic data, and models that only include demographic data are not statistically significant by observing the results of Hypothesis \ref{hyp:2} in Table \ref{tab:h2_ap}. In most cases, (\(p>0.05\)), proving that the advert viewing time data did not influence the prediction scores significantly, and that whatever correct predictions were made were most likely based on the coefficients and weights of the demographic data. Finally, observing Hypothesis \ref{hyp:3} in Table \ref{tab:h3_ap} also confirms the difference and increase of performance between advert viewing time models and those that combine advert data with demographic data. 

The exception to this rule is in category 1, where customers consistently answered "NO" in their purchase recall or purchase intention questions of the survey both in January 2017 and March 2017. Subsequently, this also influences category 5 results. Predictions seem to be high across all models when the customer has a negative purchase behavior. However, the \(t\)-test results for Hypotheses \ref{hyp:1} and \ref{hyp:3} show us that there is not a statistically significant difference between demographic data models and advert viewing time models. Because of these results, the factors that are influencing negative purchase decisions cannot be determined to be either advert viewing time or otherwise.

With these results in mind, it could be said that TV adverts are not a main factor in predicting whether a customer will change their purchasing behavior or not in a positive way, specially their actual purchase behavior. While the research based on the mere exposure effect would suggest otherwise, customers are observed to decide on their purchase without much predictability, except for their demographic data. It could be said that while there might be influence in the customer's knowledge of the brand, the data suggests that the amount of time exposed to TV adverts has no effect in the customers Actual Purchase behavior.

Other studies, using a controlled environment, have linked mere exposure with bias in consumer choice \cite[][]{janiszewski}. However, there is a possible explanation for these discrepancies in results. While controlled experiments show the TV adverts to their sample audience directly in most cases, in an uncontrolled environment of a customer's home, the customer is left free to ignore the advert and do something unrelated in the meanwhile \cite[][]{abernethy}. In the United Kingdom, there is a widely documented phenomenon involving TV advert timing and a surge in electricity caused by the use of electric kettles for preparing tea. This phenomenon is commonly called TV pickup, and has been documented for long \cite[][]{bunn,boait}. Similar to these cases, if the customers whose data were actively ignoring the adverts, the sample for training the prediction models would contain noise, altering the results. It stands to reason that without the influence of this active aversion would have on our learning model, it might correctly predict purchasing behavior as expected. However, this is more of a problem with the current TV advertisement model than with the methodology of this study. We will discuss this further in section \ref{disc_advert} of this paper.

\subsection{Influence of TV adverts based on primetime}
\label{disc_prime}

In our prediction model experiments, we used data from advertisement exposure during different time periods, days of the week and weekends. While we did this in order to observe differences in predictability for different time schedules available to different kinds of customers, especially during primetime television hours, we arrived to similar results for all time data configurations. We did not observe any difference in predictability based on Primetime television watching compared to other time periods. This could be interpreted as there being little influence in time periods and changes in purchasing behavior. 

\subsection{Implications for the TV advert industry}
\label{disc_advert}

Based on the low results of predictability of purchase behavior by advert exposure, it can be observed that TV adverts have a low probability of achieving their main purpose: to increase sales. As was stated in section \ref{disc_ap_pi} of this paper, there could be a large influence on this study's results from customers actively ignoring the adverts although they are being broadcast to their TVs. It is left to further discussion and research if adverts actually have the intended effect on customers when watched properly, or if this effect is not achieved anyway. In \cite[][]{fang} it is proposed that while the mere exposure of banner advertisement increases perceptual fluency, it doesn't have an effect on actual brand recognition compared to the control groups, for example. The existence or absence of influence by perceptual fluency on a customer's purchase decision hasn't been fully explored, but the consensus in the processing fluency model is that perceptual fluency influences brand judgement on some level, although it depends on the concept if the reception is positive or not \cite[][]{lee-a}. The problem with these studies and the current consensus, as has been said previously in this paper, is both that most experiments are done with relatively small sample sizes, and that there is a factor of uncertainty that comes with the physical avoidance of adverts in a customer's household.

With these things in mind, we consider both possibilities: either customers are attentive and the adverts have the expected influence in their short and long term memories in the case of repeated exposures \cite[][]{rossiter}; or the customers are inattentive of the advert and there might be some level of unconscious effect of mere exposure in their perception fluency \cite[][]{fang}. We observed however in our results that there is no effect on Actual Purchase behavior. While it may be true and out of the reach of our data that the customers would have influence in their memory, there was no link observed between the time of advert exposure and the purchase decisions. This raises a concern for the TV advert industry. Regardless of the cause of our results, the main implication of our paper is that currently, TV adverts are shown to have little to no effect on changes in Actual Purchase behavior. While thousands of billions of japanese yen are spent on TV advertisements each year \footnote{\label{dentsu}Dentsu, inc. 2017 Advertising Expenditures in Japan. Retrieved on May 2018 from \href {http://www.dentsu.com/knowledgeanddata/ad_expenditures/pdf/expenditures_2017.pdf}{\path{http://www.dentsu.com/knowledgeanddata/ad_expenditures/pdf/expenditures_2017.pdf}}}, the effects observed in this study are negligible. Because of this, changes are necessary in the current TV advertisement model.

\section{Limitations}
\label{limitations}

In comparison with previous research regarding this topic, our study presents a much larger database, a sample of 3000 users for 36 different products and the previously unavailable household television viewing data increases the possibilities for studying the effects of advert exposure more realistically. In accordance with this size of data, we used SVM and XGBoost, which are considered well-performing machine learning algorithms in this level of magnitude. However, while we propose using machine learning algorithms as an effective method, we are still limited by the calculation times for each model. Top performing and state of the art models, such as Deep Neural Networks, with their variations and advancements, have been known to be used with similar magnitudes of data or to expand upon it by using GAN (Generative Adversarial Network) \cite[][]{goodfellow-gan}, but their calculation time is far greater. Thus, Neural Networks are more appropriate for single models being trained, instead of a performance comparison of a large array of models as we did in our study.

Another limitation of this study is the nature of the prediction targets collected by survey. While a person can be asked directly in a survey whether they would purchase an item (purchase intention) or if they had already in the recent past (actual purchase), research based on online shopping has access to the actual purchase data, and to the number of times a person looks at a products description page, or searches terms related to it. Television advertisement research, by its nature, is harder to connect to the actual behavior of the customers and can only be assumed to be equal to their reported behavior. There is also a limitation of the number of questions that a person might answer, and how honestly they might answer them with a survey of this magnitude.

In addition to this, because of the timing of the surveys being 3 months appart between January and March 2017, we can only examine the short-term effect of advertisements, and not the long-term effect across different years of constant advertisement exposure.

Furthermore, much of the data that could be used to inspect this matter further belongs to private institutions and in many cases, is treated as a company secret. 

However, with the measurements of short-term effects of advertisement in a field where not much new research is done, we can start to shed light on problems that could be having a large impact on the costs of many industries.

\section{Conclusion and future work}
\label{conclusion}

In this paper we analyzed the ability to predict purchasing behavior, namely Actual Purchase and Purchase Intention, based on the customers' time spent exposed to television adverts using machine learning algorithms, and compared it to the ability to predict the same behavior by using demographic data on its own and in combination with the exposure time data. Based on the low prediction results of Actual Purchase by exposure time models and the relatively high prediction results for demographic based models, as well as a non-significant difference between the demographic models and the combined models, we concluded that advertisement exposure has little to no effect in short-time Actual Purchase behavior. 

We discussed possible influence by deliberate avoidance of advert cuts to prepare food or tea, and while some studies focus on the effect of attentive watching of adverts, other studies focus on the mere exposure effects, which would be achieved despite physical avoidance because of advert audio and simple proximity of the television. Both scenarios are in strong contrast with the results of our study, which shows little to no predictability in purchase behavior. Points left to research in future work are a deeper analysis of the predictable customers, looking for similarities or clusters within this class, as well as using newer and better performing deep learning algorithms when larger datasets are available. 

\section{Acknowledgements}

This paper was made possible by the data provided by Nomura Research Institute, Ltd. for their yearly Marketing Analysis Contest.

Declarations of interest: none

\section*{References}

\bibliography{ipmbibfile}

\clearpage
\appendix
\appendixpage

\section{Input data details}
\label{appendix:inputs}

As explained in section \ref{advert_viewtime}, two data configurations were used for advert viewing time. The detailed features used are shown in Table \ref{tab:viewtimes}. 

Similarly, as explained in section \ref{demographics}, demographic data was used as input for predictions in a number of our experiments. The detailed features used in the input vectors are described in Table \ref{tab:demographics}.

\begin{table}[htp] \centering
\caption{Viewing time analysis elements.}\label{tab:viewtimes}
\rowcolors{2}{DarkLime}{LightGreen}
\begin{tabular}{|m{11em}|l|}  \arrayrulecolor{white} \hline
\rowcolor{DeepGreen}
\color{white}\textbf{Data Configuration} & \color{white}\textbf{\begin{tabular}[c]{@{}l@{}}Advert Viewing Time in seconds \\ Data Features \end{tabular}} \\ \hline
\textbf{Weekdays}           
    & \begin{tabular}[c]{@{}l@{}}
    $\bullet$ Monday\\ 
    $\bullet$ Tuesday\\ 
    $\bullet$ Wednesday\\ 
    $\bullet$ Thursday\\ 
    $\bullet$ Friday\\ 
    $\bullet$ Saturday\\ 
    $\bullet$ Sunday\end{tabular} \\ \hline
\textbf{Weekday Time Slots} 
    & \begin{tabular}[c]{@{}l@{}}
    $\bullet$ Monday Primetime\\ 
    $\bullet$ Monday Non-Primetime\\ 
    $\bullet$ Tuesday Primetime\\ 
    $\bullet$ Tuesday Non-Primetime\\ 
    $\bullet$ Wednesday Primetime\\ 
    $\bullet$ Wednesday Non-Primetime\\ 
    $\bullet$ Thursday Primetime\\ 
    $\bullet$ Thursday Non-Primetime\\ 
    $\bullet$ Friday Primetime\\ 
    $\bullet$ Friday Non-Primetime\\ 
    $\bullet$ Saturday Primetime\\ 
    $\bullet$ Saturday Non-Primetime\\ 
    $\bullet$ Sunday Primetime\\ 
    $\bullet$ Sunday Non-Primetime \end{tabular} \\ \hline
\end{tabular}
\end{table}

\begin{table}[htp] \centering
\caption{Demographic data used in input vectors.}\label{tab:demographics}
\rowcolors{2}{DarkLiliac}{Liliac}
\begin{tabular}{|l|l|} \arrayrulecolor{white} \hline
\rowcolor{DeepPurple}
\color{white}\textbf{Survey Data}     & \color{white}\textbf{Possible Answers} \\ \hline
\textbf{Age}
    & \begin{tabular}[c]{@{}l@{}}
    $\bullet$ 18 to 25 years old\\ 
    $\bullet$ 26 to 35 years old\\ 
    $\bullet$ 36 to 45 years old\\ 
    $\bullet$ 46 to 55 years old\\ 
    $\bullet$ 56 or older\end{tabular} \\ \hline
\textbf{Sex}
    & \begin{tabular}[c]{@{}l@{}}
    $\bullet$ Male\\ 
    $\bullet$ Female\end{tabular} \\ \hline
\textbf{Marital Status}
    & \begin{tabular}[c]{@{}l@{}}
    $\bullet$ Single\\ 
    $\bullet$ Married\\ 
    $\bullet$ Divorced or Widowed\end{tabular} \\ \hline
\textbf{Parental Status}
    & \begin{tabular}[c]{@{}l@{}}
    $\bullet$ Parent\\ 
    $\bullet$ Not a Parent\end{tabular} \\ \hline
\textbf{Income Bracket}
    & \begin{tabular}[c]{@{}l@{}}
    $\bullet$ Not disclosed\\ 
    $\bullet$ No Income\\ 
    $\bullet$ Under 1,000,000 yen\\ 
    $\bullet$ From 1,000,000 yen to 2,000,000 yen\\ 
    $\bullet$ From 2,000,000 yen to 3,000,000 yen\\ 
    $\bullet$ From 3,000,000 yen to 4,000,000 yen\\ 
    $\bullet$ From 4,000,000 yen to 5,000,000 yen\\ 
    $\bullet$ From 5,000,000 yen to 6,000,000 yen\\ 
    $\bullet$ From 6,000,000 yen to 7,000,000 yen\\ 
    $\bullet$ From 7,000,000 yen to 10,000,000 yen\\ 
    $\bullet$ From 10,000,000 yen to 15,000,000 yen\\ 
    $\bullet$ From 15,000,000 yen to 20,000,000 yen\\ 
    $\bullet$ Over 20,000,000 yen\end{tabular} \\ \hline
\end{tabular}
\end{table}

\section{Data distribution}
\label{data_dist}

In this section we will describe the data we received from the Nomura Research Institute, Ltd., and the distribution and nature of products, adverts and prediction targets.

\subsection{Surveyed products}
\label{product_data}

\begin{figure}[htp]
\centering
\includegraphics[width=35em]{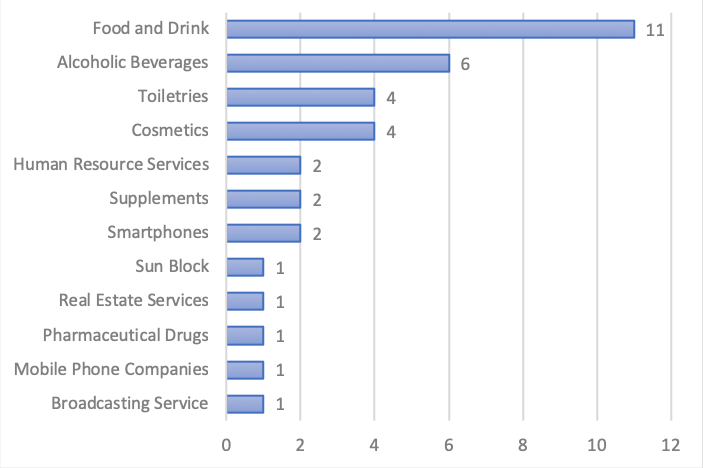}
\caption{Products matched with advert viewing data.}
\label{fig:product_dist}
\end{figure}

The surveys of purchase behavior taken in January 2017 and March 2017 included 200 products, from which only 36 were matched to television adverts during the period between both surveys. Because most of the products are sold only in Japan, a general description of their nature and distribution is explained in Figure \ref{fig:product_dist}. 

\subsection{Advert exposure and broadcasting data}
\label{advert_broadcast_dist}

\begin{figure}[htp]
\centering
\includegraphics[width=35em]{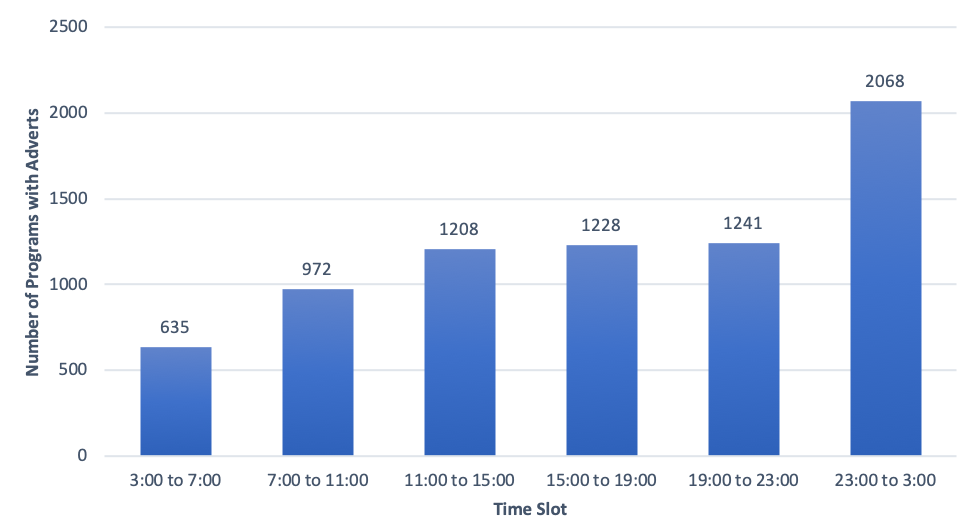}
\caption{Programs including adverts broadcast time distribution}
\label{fig:broadcast_dist}
\end{figure}

\begin{figure}[htp]
\centering
\includegraphics[width=35em]{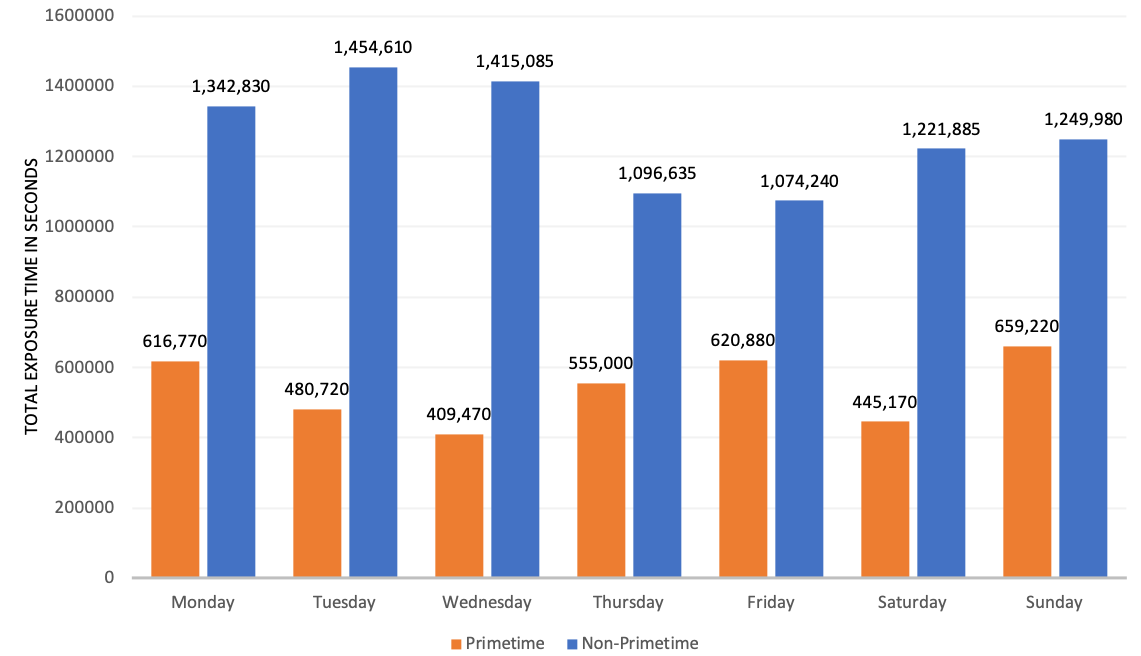}
\caption{Advert Exposure Time for all users and products by Weekday and Time Slot}
\label{fig:exposure_dist}
\end{figure}

The data we received from the Nomura Research Institute, Ltd. included the surveyees' household television viewing times and the program that was displayed when television was on. By matching this data with the adverts that were in between breaks from those programs for the products that were surveyed, we obtained the advert exposure time for each user for each product. In this study we explore the possibility of there being some difference in effect depending on the time slot, particularly the Primetime (19:00 to 23:00) time slot. In Figure \ref{fig:broadcast_dist} we show the broadcasting time distribution for the programs that displayed these 36 products during the period of time in between the survey in January 2017 and the survey in March 2017. In Figure \ref{fig:exposure_dist} we show the total sum of exposure time in seconds across all users and products for the Primetime and otherwise time slots for each day of the week.

\subsection{Prediction target categories}
\label{cat_dist}

The surveys included data of purchase intention and actual purchase at the times of January 2017 and March 2017 for 200 products, 36 of which were matched with television advert viewing data. As was explained in section \ref{target_data_cat} and in Tables \ref{tab:categories_ap} and \ref{tab:categories_pi}, we divided the data in 6 categories (0 to 5) in order to observe the changes in time for these purchase behaviors. The distributions of these categories are shown in Table \ref{tab:cat_dist}. Note that categories 4 and 5, by their nature, are a sum of categories 2 and 3, and 0 and 1 respectively. 

\begin{table}[p] \centering
\caption{Prediction target categories distribution.}\label{tab:cat_dist}
\begin{tabular}{|>{\columncolor{SweetPlum}}=c|>{\columncolor{Salmon}}+r|>{\columncolor{Liliac}}+r|>{\columncolor{Salmon}}+r|>{\columncolor{Liliac}}+r|} \arrayrulecolor{white} \hline
\rowcolor{Plum}

\cellcolor{white}{}    & \multicolumn{2}{c|}{\color{white}\textbf{All products (200)}} & \multicolumn{2}{c|}{\begin{tabular}[c]{@{}c@{}}\color{white}\textbf{Advert matched} \\ \color{white}\textbf{products (36)}\end{tabular}} \\ \hline
\rowstyle{\bfseries}
\cellcolor{Plum}\color{white}Category
    & \cellcolor{DarkSalmon}\begin{tabular}[c]{@{}c@{}}Actual \\ Purchase \end{tabular} 
    & \cellcolor{DarkLiliac}\begin{tabular}[c]{@{}c@{}}Purchase \\ Intention \end{tabular} 
    & \cellcolor{DarkSalmon}\begin{tabular}[c]{@{}c@{}}Actual \\ Purchase \end{tabular}
    & \cellcolor{DarkLiliac}\begin{tabular}[c]{@{}c@{}}Purchase \\ Intention \end{tabular} \\ \hline

    0   & 6\%     & 8\%     & 6\%     & 8\%     \\ \hline
    1   & 73\%    & 59\%    & 76\%    & 58\%    \\ \hline
    2   & 8\%     & 9\%     & 7\%     & 8\%     \\ \hline
    3   & 13\%    & 24\%    & 10\%    & 26\%    \\ \hline
    4   & 21\%    & 33\%    & 17\%    & 35\%    \\ \hline
    5   & 79\%    & 67\%    & 83\%    & 65\%    \\ \hline                    
\end{tabular}
\end{table}

\end{document}